\documentclass[reqno,12pt]{amsart}
\usepackage{amssymb,amsmath}
\usepackage{bbm}
\usepackage{fullpage}

\begin{document}

\def\sech{\mathrm{sech}}
\def\csch{\mathrm{csch}}
\def\Re{{\rm Re}\,}
\def\Im{{\rm Im}\,}
\def\sgn{{\rm sgn}}
\def\arg{{\rm arg}}

\newtheorem{prop}{Proposition}
\newtheorem{thm}{Theorem}
\newtheorem{cor}{Corollary}

\newenvironment{changemargin}[2]{%
  \begin{list}{}{%
    \setlength{\topsep}{0pt}%
    \setlength{\leftmargin}{#1}%
    \setlength{\rightmargin}{#2}%
    \setlength{\listparindent}{\parindent}%
    \setlength{\itemindent}{\parindent}%
    \setlength{\parsep}{\parskip}%
  }%
  \item[]}{\end{list}}

\title{Travelling waves and conservation laws for complex mKdV-type equations}

\author{
Stephen C. Anco, Mohammad Mohiuddin, Thomas Wolf\\
Department of Mathematics\\
Brock University}

\begin{abstract}
Travelling waves and conservation laws are studied 
for a wide class of $U(1)$-invariant complex mKdV equations 
containing the two known integrable generalizations of 
the ordinary (real) mKdV equation. 
The main results on travelling waves include deriving 
new complex solitary waves and kinks that generalize 
the well-known mKdV $\sech$ and $\tanh$ solutions. 
The main results on conservation laws consist of explicitly finding 
all 1st order conserved densities that yield phase-invariant counterparts of 
the well-known mKdV conserved densities for 
momentum, energy, and Galilean energy, 
and a new conserved density describing 
the angular twist of complex kink solutions. 
\end{abstract}

\maketitle

\section{Introduction}

In this paper, 
we study travelling wave solutions and conservation laws of general complex mKdV-type equations
\begin{equation}\label{mkdveqn}
u_t+\alpha\bar{u}uu_x+\beta u^2\bar{u}_x+\gamma u_{xxx}=0 
\end{equation}
for $u(t,x)$ with complex coefficients 
$\alpha=\alpha_1+i\alpha_2$, $\beta=\beta_1+i\beta_2$,
and real coefficient $\gamma>0$. 
An equation of the form (\ref{mkdveqn}) is equivalent to a coupled nonlinear system 
\begin{align}
\begin{split}
&u_{1t}+((\alpha_1+\beta_1)u_1^2-2\beta_2u_1u_2+(\alpha_1-\beta_1)u_2^2)u_{1x}\\
&\qquad -((\alpha_2-\beta_2)u_1^2-2\beta_1u_1u_2+(\alpha_2+\beta_2)u_2^2)u_{2x}+\gamma u_{1xxx}=0
\end{split}\label{u1veqn}\\
\begin{split}
&u_{2t}+((\alpha_2+\beta_2)u_1^2+2\beta_1u_1u_2+(\alpha_2-\beta_2)u_2^2)u_{1x}\\
&\qquad +((\alpha_1-\beta_1)u_1^2+2\beta_2u_1u_2+(\alpha_1+\beta_1)u_2^2)u_{2x}
+\gamma u_{2xxx}=0
\end{split}\label{u2veqn}   
\end{align}
for the real and imaginary parts of $u(t,x)=u_1(t,x)+iu_2(t,x)$.
This system reduces to the ordinary mKdV equation in the case when 
$u_2=0$ and $\alpha_2=\beta_2=0$. 

Complex mKdV-type equations \eqref{mkdveqn} 
are interesting both physically and mathematically.
When the coefficients $\alpha$ and $\beta$ are real, 
such equations describe propagation of short pulses in optical fibers 
\cite{RadLak,GilHieNimOht}
where the physical meaning of $t$ and $x$ as time and space variables is reversed. 
In the cases where the ratio of these real coefficients is $\beta/\alpha=0$ or $\beta/\alpha=1/3$, 
the resulting equations are integrable systems \cite{Hir1973,SasSat1991},
possessing rich mathematical features such as soliton solutions 
which describe nonlinear interactions of two or more travelling waves, 
as well as a hierarchy of conservation laws which involve 
$u,\bar{u},u_x,\bar{u}_x$, and increasingly higher order $x$-derivatives of $u$ and $\bar{u}$. 
In contrast, 
when the coefficients $\alpha$ and $\beta$ are complex, 
little seems to be known about the nature of solutions 
or the existence of any integrability structure 
for complex mKdV-type equations \eqref{mkdveqn}.
The equivalent coupled systems \eqref{u1veqn}--\eqref{u2veqn}   
with $\alpha_2\neq 0$ or $\beta_2\neq 0$ 
arise in modelling weakly coupled two-layer fluids \cite{GaoTan}. 

Travelling waves for the class of complex mKdV-type equations (\ref{mkdveqn}) 
are considered in section \ref{travellingwaves}. 
By use of symmetry reduction and integrating factors, 
we first derive all smooth solutions of the form 
\begin{equation}\label{travellingwave}
u(t,x) = U(x-ct)= a+bf(x-ct)
\end{equation}
in which $a=a_1+ia_2, b=b_1+ib_2$ are complex constants, and $f(x)$ is a real valued function that
either is single-peaked and vanishes for large $x$ (i.e.\ a solitary wave), 
or has no peak and approaches different constant values for large positive/negative $x$ (i.e.\ a kink).
We next derive all smooth solutions having a linear phase 
\begin{equation}\label{LINEARPHASE}
u(t,x)=\exp(i(kx+wt+\phi))f(x-ct)
\end{equation} 
where $k,w,\phi$ are real constants, and $f(x)$ is again a real valued function with the same
general profile (i.e. either a solitary wave or a kink) as considered for the solutions (\ref{travellingwave}).
Both of these two classes of complex travelling waves (\ref{travellingwave}) and (\ref{LINEARPHASE}) 
include the familiar $\sech$ and $\tanh$ profiles for $f(x)$ 
in the case of the ordinary mKdV equation \cite{Hir1972}. 

Conservation laws for the class of complex mKdV-type equations (\ref{mkdveqn}) 
are then considered in section \ref{conlaws}. 
Specifically, by means of multipliers, 
we derive all conserved densities and fluxes of the form 
\begin{equation}\label{2ndorderT}
T(t,x,u,\bar{u},u_x,\bar{u}_x,u_{xx},\bar{u}_{xx})
\end{equation}
\begin{equation}\label{2ndorderX}
X(t,x,u,\bar{u},u_x,\bar{u}_x,u_{xx},\bar{u}_{xx},u_{xxx},\bar{u}_{xxx},u_{xxxx},\bar{u}_{xxxx})
\end{equation}
which satisfy
\begin{equation}\label{conslaw}
D_t T+D_x X = 0
\end{equation}
corresponding to conserved quantities
\begin{equation}\label{CQuantities}
C=\int_{-\infty}^\infty T\,dx = \text{const.}
\end{equation}
for all solutions $u(t,x)$ that have vanishing flux at $x = \pm \infty$. 
The class of conserved densities (\ref{2ndorderT}) includes the well-known conservation laws 
for mass, momentum, energy and Galilean energy in the case of the ordinary mKdV equation \cite{MiuGarKru},
as well as the first of the higher-derivative conservation laws 
in the hierarchy arising for the two known integrable cases of complex mKdV equations.

In section \ref{solnfeatures}, 
the conserved quantities (\ref{CQuantities}) are used to explore 
some features of the travelling wave solutions (\ref{travellingwave}) and (\ref{LINEARPHASE}).
Finally, a summary of the main new results obtained in the previous sections
is provided in section \ref{summarize}. 

Hereafter, for convenience we will put 
\begin{equation}
\gamma=1
\end{equation}
by scaling the time and space variables 
$t\rightarrow \sqrt {\gamma} t$, $x\rightarrow \sqrt {\gamma} x$.

\section{Travelling waves}\label{travellingwaves}

Each complex mKdV equation in the class (\ref{mkdveqn}) has the following basic invariance properties:
\begin{align}
&\text{scaling}\qquad & 
x\rightarrow \lambda x, t\rightarrow \lambda ^3 t, u\rightarrow \lambda ^{-1} u \label{scaling}\\
&\text{time translation}\qquad  & 
t\rightarrow t+\epsilon 
\label{timetrans}\\
&\text{space translation}\qquad &
x\rightarrow x+\epsilon 
\label{spacetrans}\\
&\text{phase rotation}\qquad &
u\rightarrow \exp(i\phi)u
\label{phaserot}
\end{align}
where the parameters are given by 
$\lambda \neq 0$, $-\infty<\epsilon <\infty$, $0\leq\phi< 2\pi$.
Composition of these transformations (\ref{scaling})--(\ref{phaserot}) 
yields a 4-parameter group of point symmetries admitted by all equations (\ref{mkdveqn}).

Travelling waves $u = U(x-ct)$ arise naturally as group-invariant solutions 
\cite{1stbook} with respect to the combined space-time translation 
\begin{equation}\label{sttranslation}
t\rightarrow t+\epsilon , x\rightarrow x+c\epsilon
\end{equation}
where $c$ is the speed of the wave. 
From equation (\ref{mkdveqn}), such solutions satisfy 
\begin{equation}\label{UODE}
-cU'+\alpha \bar{U}UU' + \beta U^2\bar{U}' + U'''=0     
\end{equation}
which is a 3rd order, nonlinear, complex ODE for $U(\xi)$, in terms of the invariant variable
\begin{equation}\label{InvariantVariable}
\xi = x-ct .
\end{equation}
Note this ODE (\ref{UODE}) inherits invariance under phase rotations
\begin{equation}\label{Protation}
U\rightarrow \exp(i\phi)U .
\end{equation}
When the coefficients $\alpha$ and $\beta$ are real, 
we can find all real solutions $U(\xi)$ straightforwardly 
by the use of integrating factors. 
In contrast, 
when the coefficients are complex or when we seek all complex solutions, 
the ODE (\ref{UODE}) cannot in general be integrated explicitly. 
However, the class of solutions 
\begin{equation}\label{travellingwave1}
U= a+bf(\xi)
\end{equation}
given by a real function $f(\xi)$ and complex constants $a$ and $b$ will by-pass
these difficulties, as we now show.

\subsection{Solitary waves and Kinks}\label{SolitaryKinks}

For travelling wave solutions of the form (\ref{travellingwave1}), 
the ODE (\ref{UODE}) is given by 
\begin{equation}\label{fODEE}
0=(\alpha \bar{a}ab+\beta a^2\bar{b}-cb)f'+((\alpha+2\beta)ab\bar{b}+\alpha \bar{a}b^2)ff'
+(\alpha+\beta)b^2\bar{b}f^2f'+bf'''
\end{equation}
where, under phase rotations (\ref{Protation}), 
\begin{equation}\label{ABrotation}
a\rightarrow \exp(i\phi)a, b\rightarrow \exp(i\phi)b, f\rightarrow f .
\end{equation}
Consequently, 
the ODE (\ref{fODEE}) for $f(\xi)$ can be simplified 
by using a phase rotation (\ref{ABrotation}) to put $b=|b|$, 
which then gives 
\begin{equation}\label{ffODEE}
0 = (A+Bf+Cf^2)f'+f'''
\end{equation}
with coefficients
\begin{equation}
\begin{aligned}
&
A = \alpha |a|^2 +\beta a^2 -c ,
\\
&
B = (\alpha (a+\bar{a})+2\beta a) |b|,
\\
&
C = (\alpha +\beta) |b|^2 .
\end{aligned}
\end{equation}
Since $f$ is real, these coefficients must be real, 
and hence we require the conditions
\begin{equation}\label{E:first}
\Im A = \Im B = \Im C =0 .
\end{equation}
Solving these conditions, we obtain two cases:
\begin{equation}\label{E:first1}
a_2 \neq 0, \quad \beta_1 = \alpha_2=\beta_2 =0 ;
\end{equation}
\begin{equation}\label{E:first2}
a_2=0, \quad \alpha_2+\beta_2 =0 .
\end{equation}
For each case, 
the ODE (\ref{ffODEE}) is straightforward to solve by integrating factors.

We will be interested first in solutions $f(\xi)$ 
that describe a solitary wave of the form (\ref{travellingwave}) 
having a single peak at $\xi=\xi_0$ 
and decaying to a constant value for $|\xi|\rightarrow \infty$. 
Without loss of generality, 
these properties will hold if $f(\xi)$ satisfies the conditions 
(i) 
$f\rightarrow 0$ as $\xi\rightarrow \pm\infty$ 
and (ii) 
$f' = 0, f'' \neq 0$ at $\xi = \xi_0$.
Condition (i) provides asymptotic boundary conditions
\begin{equation}\label{Bcondision1}
f(\pm \infty) = f'(\pm \infty) = f''(\pm \infty) = 0
\end{equation}
which we impose on the solution of ODE (\ref{ffODEE}). 
Integrating this ODE once, we get
\begin{equation}\label{Bcondision2}
0 = Af + \tfrac{1}{2}Bf^2 + \tfrac{1}{3}Cf^3 +f''
\end{equation} 
after the asymptotic boundary conditions (\ref{Bcondision1}) are imposed.
Then using the integrating factor $f'$ 
and again imposing the asymptotic boundary conditions (\ref{Bcondision1}), 
we obtain
\begin{equation}\label{Bcondision3}
0 = Af^2+\tfrac{1}{3}Bf^3 + \tfrac {1}{6}Cf^4+f'^2
\end{equation} 
which is a separable 1st order ODE. 
The general solution satisfying condition (ii) is given by
\begin{equation}\label{fsolitarywave}
f(\xi) =
\begin{cases}
\displaystyle
\frac{-6A}{\sqrt{B^2-6AC} \cosh(\sqrt{-A}(\xi-\xi_0)) +B}, 
& A\neq 0, B\neq 0 \; \text{or} \; C \neq 0 \\\\
\displaystyle
\exp(\pm\sqrt{-A}(\xi-\xi_0)), 
& A \neq 0, B = C =0\\\\
\displaystyle
\frac{-2B}{(B^2/6)(\xi-\xi_0)^2 +C}, 
&  A= 0, B \neq 0\\\\
\displaystyle
\pm\frac{\sqrt{-6/C}}{\xi-\xi_0}, 
& A = B = 0, C \neq 0
\end{cases}
\end{equation}
with the coefficients $A,B,C$ given by the two cases (\ref{E:first1}) and (\ref{E:first2}).
This leads to the following classification result.

\begin{prop}\label{fsolns1}
Modulo phase rotations 
$a\rightarrow \exp(i\phi)a$, $b\rightarrow \exp(i\phi)b$, 
and translations 
$\xi\rightarrow \xi+\xi_0$,
the travelling wave ODE (\ref{fODEE}) with asymptotic boundary conditions (\ref{Bcondision1})
has six real piecewise-smooth solutions:
\begin{subequations}\label{solutions1}
\begin{align}
& f = -\frac{12 a/|b|}{2c\xi^2+3}\\
\intertext{with}
&\alpha_2 + \beta_2 = 0, \alpha_1 + \beta_1 \neq 0\\
& a = \sqrt{\frac{c}{\alpha_1 + \beta_1}}, \arg\ b = 0 ;
\end{align}
\end{subequations}
\begin{subequations}\label{solutions2}
\begin{align}
&f = \frac{6\csch(\Theta) a/|b|}{\sqrt{6+4\sinh^2(\Theta)}\cosh(\sech(\Theta)\sqrt{c}\xi)+2\sinh(\Theta)}\\
\intertext{with}
&\alpha_2 + \beta_2 = 0, \alpha_1 + \beta_1 \neq 0\\
& a= \sqrt{\frac{c}{\alpha_1+\beta_1}}\tanh(\Theta), \arg\ b = 0\\
\intertext{where $-\infty<\Theta<\infty$;}\nonumber
\end{align}
\end{subequations}
\begin{subequations}\label{solutions3}
\begin{align}
&f = \exp(-\sqrt{c}|\xi|)/|b|\\
\intertext{with}
&\alpha_2 +\beta_2 = \alpha_1 +\beta_1 = 0, \alpha_1 \neq 0 \; \text{or} \; \alpha_2 \neq 0\\
& a = \sinh(\Theta), \arg\ b =0\\
\intertext{where $-\infty<\Theta<\infty$;}\nonumber
\end{align}
\end{subequations}
\begin{subequations}\label{solutions4}
\begin{align}
&f = -\frac{12 \Re a/|b|}{2c \cos^2(\theta)\xi^2+3}\\
\intertext{with}
&\alpha_2 = \beta_1= \beta_2 = 0, \alpha_1 \neq 0\\
& a = \sqrt{\frac{c}{\alpha_1}}\exp(i\theta) ,\arg\ b=0\\
\intertext{where $0\leq\theta < 2\pi$;}\nonumber
\end{align}
\end{subequations}
\begin{subequations}\label{solutions5}
\begin{align}
&f =\frac{6\csch(\Theta) |a|/|b|}{\sqrt{6+4\cos^2(\theta)\sinh^2(\Theta)}\cosh(\sech(\Theta)\sqrt{c}\xi)+2\cos(\theta)\sinh(\Theta)}\\
\intertext{with}
&\alpha_2 = \beta_1= \beta_2 = 0, \alpha_1 \neq 0\\
& a= \sqrt{\frac{c}{\alpha_1}}\tanh(\Theta)\exp(i\theta) ,\arg\ b=0\\
\intertext{where $-\infty<\Theta<\infty$ and $0\leq \theta <2\pi$;}\nonumber
\nonumber
\end{align}
\end{subequations}
\begin{subequations}\label{solutions6}
\begin{align}
&f = \pm \frac{|a|/|b|}{\sqrt{-c/6}\xi}\\
\intertext{with}
&\alpha_2 = \beta_1 = \beta_2 = 0, \alpha_1 \neq 0\\
& a = i\sqrt{\frac{c}{\alpha_1}} ,\arg\ b=0.
\end{align}
\end{subequations}
\end{prop}

We will next be interested in solutions $f(\xi)$ 
that describe a kink of the form (\ref{travellingwave})
having different asymptotic values for $\xi\rightarrow \pm \infty$ 
and exhibiting an inflection at $\xi = \xi_0$.
Without loss of generality, 
these properties will hold if $f(\xi)$ satisfies the conditions
(i) 
$f\rightarrow f_\pm = \text{const.}$ 
with $f_+ \neq f_-$ as $\xi\rightarrow \pm \infty$ 
and (ii) 
$f'' = 0$ at $\xi = \xi_0$. 
Since the ODE (\ref{ffODEE}) for $f(\xi)$ is invariant under shifts
$f\rightarrow f-\epsilon$, $a\rightarrow a+b\epsilon$, $b\rightarrow b$, 
we can put $f_+ = -f_- = f_0 \neq 0$
by shifting $f\rightarrow f-(f_+ + f_-)/2$.
Hence condition (i) imposes asymptotic boundary conditions
\begin{equation}\label{Abcondition}
f(+\infty) = -f(-\infty) = f_0, \quad f'(\pm \infty) = f''(\pm \infty) = 0 .
\end{equation}
Now, integration of ODE (\ref{ffODEE}) by means of the same integrating factors used earlier 
yields the separable 1st order ODE 
\begin{equation}\label{ODE3}
0 = D + Ef + Af^2 + \tfrac{1}{3}Bf^3 +\tfrac{1}{6}Cf^4 +f'^2
\end{equation}
where $D$ and $E$ are constants. 
The easiest way we can impose the asymptotic boundary conditions (\ref{Abcondition}) 
is by considering the integral curves of this ODE (\ref{ODE3}) 
in the phase plane given by 
$f' = (-D - Ef - Af^2 - \tfrac{1}{3}Bf^3 -\tfrac{1}{6}Cf^4 )^{1/2}$ 
as a function of $f$, 
where these conditions can be expressed as 
$f'|_{f=\pm f_0} = 0$ and $\lim_{f \to \pm f_0} \int (1/f')df = \pm \infty$.
For these phase plane conditions to hold, we must have
\begin{equation}
\frac{6}{C}(D + Ef + Af^2 + \tfrac{1}{3}Bf^3 +\tfrac{1}{6}Cf^4) 
=  (f-f_0)^2 (f+f_0)^2
\end{equation}
which requires 
\begin{equation}\label{CCond4}
B=E=0, D=\tfrac{3}{2}A^2/C
\end{equation}
and
\begin{equation}\label{CCond5}
f_0 = \sqrt{-3A/C} .
\end{equation}
Combining conditions (\ref{CCond4}) and conditions (\ref{E:first}) on the coefficients,
we obtain two cases:
\begin{subequations}
\begin{equation}\label{kinkcase1}
a_1 =0, a_2 \neq 0, \quad\beta_1=\alpha_2 = \beta_2 = 0 ;
\end{equation}
\begin{equation}\label{kinkcase2}
a_1 = a_2 = 0, \quad \alpha_2 + \beta_2 = 0.
\end{equation}
\end{subequations}
In both cases, the general solution of the ODE (\ref{ffODEE}) satisfying condition (ii)
is given by
\begin{equation}\label{genfsoln2}
f(\xi)= \sqrt{-3A/C} \tanh(\sqrt{A/2}\xi), \quad A \neq 0, C \neq 0.
\end{equation}
This leads to the following classification result.

\begin{prop}\label{fsolns2}
Modulo phase rotations and shifts 
$a\rightarrow \exp(i\phi)(a+b\epsilon)$, $b\rightarrow \exp(i\phi)b$,
and translations $\xi\rightarrow \xi+\xi_0$, 
the travelling wave ODE (\ref{fODEE}) 
with asymptotic boundary conditions (\ref{Abcondition}) 
has two real piecewise-smooth solutions:
\begin{subequations}\label{Rsolutions1}
\begin{align}
&f = \sqrt{\frac{3c}{\alpha}}\sech(\Theta) \tanh(\sech(\Theta)\sqrt{-c/2}\xi)/|b|\\
\intertext{with}
&\alpha_2 =\beta_1=\beta_2 =0, \alpha_1 \neq 0\\
& a = i\sqrt{\frac{c}{\alpha}}\tanh(\Theta), \arg\ b=0\\
\intertext{where $-\infty<\Theta<\infty$;}\nonumber
\end{align}
\end{subequations}
\begin{subequations}\label{Rsolutions2}
\begin{align}
&f = \sqrt{\frac{3c}{\alpha+\beta}}\tanh(\sqrt{-c/2}\xi)/|b|\\
\intertext{with}
&\alpha_2 +\beta_2 =0, \alpha_1 +\beta_1 \neq 0\\
& a = 0, \arg\ b=0 .
\end{align}
\end{subequations}
\end{prop}

The solutions obtained in Propositions \ref{fsolns1} and \ref{fsolns2}
have interesting analytical behaviour. 
Solutions (\ref{solutions2}), (\ref{solutions5}), 
(\ref{Rsolutions1}), (\ref{Rsolutions2})
are smooth for all $\xi$, 
i.e.\ $f\in C^\infty$ on $-\infty<\xi<\infty$. 
Solution (\ref{solutions6}) 
has a blow-up singularity at $\xi =0$, 
i.e.\ $|f|\rightarrow \infty$ as $\xi\rightarrow 0$. 
Solutions (\ref{solutions1}) and (\ref{solutions4}) are smooth for all $\xi$ 
iff $c>0$, since otherwise there is a blow-up singularity at some point 
$\xi = \xi_0$ (such that their denominators vanish). 
In contrast, 
solution (\ref{solutions3})
is continuous for all $\xi$ but has a cusp at $\xi =0$, 
i.e.\ $f\in C^0$ on $-\infty<\xi<\infty$
while $f\in C^\infty$ only on $0<|\xi|<\infty$
such that $f'$ is discontinuous at $\xi=0$. 

\begin{thm}\label{solitarywavesolns}
For a complex mKdV equation (\ref{mkdveqn}), 
non-singular solitary wave solutions of the form (\ref{travellingwave}) 
having a single peak at $x=ct$ and decaying to a constant value for $|x|\rightarrow \infty$ 
are admitted only in the following two cases: 
\newline
{\rm (i)} 
$\Im \alpha = 0$, $\beta = 0$
\begin{align}
&
u(t,x)=e^{i\phi}\sqrt{\frac{c}{\alpha}}\left(
e^{i\theta} \tanh\Theta+\frac{6\sech\Theta}
{2\cos\theta \sinh\Theta+\sqrt{6+4\cos^2\theta\sinh^2\Theta}\cosh(\sqrt{c}(x-ct)\sech\Theta)}
\right)
\label{solitary1}\\
& 
u(t,x)=e^{i\phi}\sqrt{\frac{c}{\alpha}}\left(
e^{i\theta}-\frac{12\cos\theta}{3+2c(x-ct)^2 \cos^2\theta}
\right)
\label{solitary2}
\end{align}
where $c>0$, $\alpha>0$, and the parameters are given by 
$-\infty<\Theta<\infty$, $0\leq\theta <2\pi$ and $0\leq \phi< 2\pi$. 
\newline
{\rm (ii)} 
$\Im(\alpha+\beta)=0$ 
\begin{align}
&u(t,x)=e^{i\phi}\sqrt{\frac{c}{\alpha+\beta}}\left(
\tanh\Theta+\frac{6\sech\Theta}{2\sinh\Theta
+\sqrt{6+4\sinh^2\Theta}\cosh(\sqrt{c}(x-ct)\sech\Theta)}
\right)
\label{solitary3}\\
&
u(t,x)=e^{i\phi}\sqrt{\frac{c}{\alpha+\beta}}\left(1-\frac{12}{3+2c(x-ct)^2}\right)
\label{solitary4}
\end{align}
where $c>0$, $\alpha+\beta>0$, and the parameters are given by 
$-\infty<\Theta<\infty$ and $0\leq\phi< 2\pi$. 
\end{thm}

\begin{thm}\label{cuspsolns}
For a complex mKdV equation (\ref{mkdveqn}), 
a solitary wave solution of the form (\ref{travellingwave}) 
with a cusp singularity at $x=ct$ 
and a non-singular decaying tail for $|x|\rightarrow \infty$ 
is admitted in the case 
$\alpha+\beta=0$: 
\begin{equation}\label{solitary5}
u(t,x)=e^{i\phi}\left(\sinh\Theta+\exp(-\sqrt{c}|x-ct|)\right)
\end{equation}
where $c>0$, and the parameters are given by 
$-\infty<\Theta<\infty$ and $0\leq \phi< 2\pi$. 
\end{thm}

Solutions (\ref{solitary1}) and (\ref{solitary3}) 
generalize the familiar solitary wave 
\begin{equation}\label{sechsoln}
u(t,x) = e^{i\phi}\frac{\sqrt{6c/(\alpha+\beta)}}{\cosh(\sqrt{c}(x-ct))}
\end{equation}
which is well known \cite{AncTchWil}
in the case of complex mKdV equations with real coefficients $\alpha=\bar{\alpha}$ and $\beta=\bar{\beta}$. 
We note this solitary wave (\ref{sechsoln}) itself is still a solution 
in the more general case of complex mKdV equations 
whose coefficients satisfy $\Im (\alpha+\beta)=0$, 
and it reduces to the 1-soliton solution in the case of the ordinary mKdV equation (where $\phi =0$).

Solution (\ref{solitary5}) is somewhat trivial because complex mKdV equations 
in the case $\alpha+\beta =0$ reduce to the Airy equation $u_t+u_{xxx}=0$ 
when $u$ has a constant phase.

\begin{thm}\label{kinksolns}
For a complex mKdV equation (\ref{mkdveqn}), 
non-singular kink solutions of the form (\ref{travellingwave}) 
having an inflection at $x=ct$ 
and approaching different constants for $x\rightarrow \pm \infty$ 
are admitted only in the following two cases:
\newline
{\rm (i)} $\Im (\alpha+\beta)=0$
\begin{equation}\label{kink1}
u(t,x)=e^{i\phi} \sqrt{\frac{3c}{\alpha+\beta}}\tanh(\sqrt{-c/2}(x-ct))
\end{equation}
where $c<0$, $\alpha+\beta<0$, and the parameter is given by 
$0\leq \phi< 2\pi$. 
\newline
{\rm (ii)} $\Im \alpha=0$, $\beta=0$
\begin{equation}\label{kink2}
u(t,x)=e^{i\phi} \sqrt{\frac{c}{\alpha}}\left(
i\tanh\Theta+\sqrt{3}\sech\Theta\tanh(\sqrt{-c/2}(x-ct)\sech\Theta)
\right)
\end{equation}
where $c<0$, $\alpha<0$, and the parameters are given by 
$-\infty<\Theta<\infty$ and $0\leq \phi< 2\pi$. 
\end{thm}

Up to a phase rotation, 
solution (\ref{kink1}) is just the familiar kink solution 
known for the ordinary mKdV equation. 
We see the same kink solution also holds 
for more general complex mKdV equations. 
Solution (\ref{kink2}) is a complex generalization of this kink, 
having asymptotic complex values
$u\rightarrow e^{i\phi} \sqrt{c/\alpha}(i\tanh\Theta\pm \sqrt{3}\sech\Theta)$ 
as $x\rightarrow \pm \infty$.

Additional features of these solitary wave solutions (\ref{solitary1})--(\ref{solitary5}) 
and kink solutions (\ref{kink1})--(\ref{kink2}) 
will be discussed in section \ref{solnfeatures}.

\subsection{Waves with a linear phase}\label{Wavesphase}

Travelling waves $u=U(x-ct)$ have a natural generalization given by 
$u=\exp(i(kx+wt+\phi))U(x-ct)$
whose form is group-invariant with respect to 
space-time translations combined with phase rotations
\begin{equation}\label{PROTATION}
t\rightarrow t+\epsilon , x\rightarrow x+c\epsilon, 
u\rightarrow \exp(i(kc+w)\epsilon )u.
\end{equation}
In this generalization the wave is modulated by a linear phase 
where $k,w$ are the inverse wave length and the frequency of the modulation. 
A complex mKdV equation (\ref{mkdveqn}) will admit such solutions 
if $U(\xi)$ satisfies the 3rd order, nonlinear, complex ODE
\begin{equation}\label{nonlinearcomplex}
0=i(w-k^3)U+i(\alpha-\beta) k\bar{U}U^2-(c+3k^2)U'+\alpha\bar{U}UU'+\beta U^2\bar{U}'+i3kU''+U'''
\end{equation}
in terms of the invariant variable (\ref{InvariantVariable}). 
This ODE (\ref{nonlinearcomplex}) cannot in general be integrated explicitly. 
Nevertheless, as we now show, 
it is possible to find a class of explicit solutions 
\begin{equation}\label{REALF}
U=\bar{U}=f(\xi)
\end{equation} 
where $f(\xi)$ is a real function. 
For solutions of this form (\ref{REALF}), 
the complex ODE (\ref{nonlinearcomplex}) splits into a pair of real ODEs
\begin{align}
&0=f'''+((\alpha_1+\beta_1)f^2-c-3k^2)f'+(\beta_2-\alpha_2)kf^3 
\label{eqns1}\\
&0=3kf''+(\alpha_2+\beta_2)f^2f'+(w-k^3)f +(\alpha_1-\beta_1)kf^3 
\label{eqns2}
\end{align}
constituting an overdetermined system for $f(\xi),k,w$.

This system (\ref{eqns1})--(\ref{eqns2}) can be solved by an integrability analysis, 
which we carry out using computer algebra, 
as summarized in section \ref{Summarize2}. 
The analysis yields all of the different cases in which the system 
reduces to a single ODE for $f(\xi)$ plus algebraic conditions on $k,w$.

\begin{prop}\label{solslinearphase}
All solutions of the ODE system (\ref{eqns1})--(\ref{eqns2}) for travelling waves
with a non-zero linear phase are given by:
\begin{subequations}
\begin{align}
&f''-(c+3k^2)f+\tfrac{1}{3}\alpha_1 f^3=0 \\
\intertext{with} 
&\alpha_2 =\beta_1=\beta_2 =0, \alpha_1 \neq 0 \\
&w=-(3c+8k^2)k ;
\end{align}\label{fequations1}
\end{subequations}
\begin{subequations}
\begin{align}
&f' \mp\sqrt{c+3k^2}f=0 \\
\intertext{with} 
&\alpha_1^2 +\alpha_2^2=\beta_1^2 +\beta_2^2 \neq 0 \\
&w=-(3c+8k^2)k \\
&\pm\sqrt{c+3k^2}(\alpha_1+\beta_1)-(\alpha_2-\beta_2)k=\pm\sqrt{c+3k^2}(\alpha_2+\beta_2)+(\alpha_1-\beta_1)k=0. 
\end{align}\label{fequations2}
\end{subequations}
\end{prop}

We will be interested in solutions such that $f(\xi)$ has the profile of either a solitary wave or 
a kink, as described by the respective asymptotic boundary conditions (\ref{Bcondision1}) or (\ref{Abcondition}).
Integrating the ODEs (\ref{fequations1}) and  (\ref{fequations2}) under these boundary conditions, we obtain the 
following classification results.

\begin{thm}\label{linphasesolitarywaves}
For a complex mKdV equation (\ref{mkdveqn}), 
non-singular solitary wave solutions 
having a non-zero linear phase of the form (\ref{LINEARPHASE}) 
are admitted only in the case
$\Im \alpha =0$, $\beta=0$:
\begin{equation}\label{SWsolutions1}
u(t,x)=
\sqrt{\frac{6(c+3k^2)}{\alpha}}\exp(i\phi)
\frac{\exp(ik(x-(3c+8k^2)t))}{\cosh((x-ct)\sqrt{c+3k^2})}
\end{equation}
where $c>-3k^2$, $\alpha>0$, and the parameters are given by 
$-\infty<k<\infty$, $0\leq\phi<2\pi$. 
\end{thm}

\begin{thm}\label{linphasepeakon}
A solitary wave solution of the form (\ref{LINEARPHASE}) 
with a cusp singularity at $x=ct$ and a non-singular decaying tail 
for $|x|\rightarrow \infty$ is admitted in the case 
$|\alpha|=|\beta|$:
\begin{equation}\label{SWsolutions3}
u(t,x)=
A\exp\big(i\phi\big)\exp\big(-i\epsilon \sqrt{\frac{c}{\sigma^2-3}}\big(x-\frac{3\sigma^2-1}{\sigma^2-3}ct\big)\big)
\exp\big(-\sqrt{\frac{c\sigma^2}{\sigma^2-3}}|x-ct|\big)
\end{equation}
with $\sigma=\Re (\alpha-\beta)/\Im (\alpha+\beta)=-\Im (\alpha-\beta)/\Re (\alpha+\beta) \neq 0$ 
and 
$\epsilon =\sgn(\sigma(x-ct))=\pm 1$, 
where $c>0$, $\sigma^2>3$, or $c<0$, $\sigma^2<3$,
and the parameters are given by $A>0$, $0\leq\phi<2\pi$. 
\end{thm}

Solution (\ref{SWsolutions1}) is the well-known 1-soliton solution 
for the integrable case $\alpha=\bar{\alpha}$, $\beta=0$ 
in the class of complex mKdV equations (\ref{mkdveqn}).
It reduces to the familiar solitary wave solution (\ref{sechsoln}) when $k=0$.

In contrast, solution (\ref{SWsolutions3}) is new. 
By writing it in terms of $\xi=x-ct$ and $t$, 
we see that its amplitude 
$|u|= A\exp(-\sqrt{c\sigma^2/(\sigma^2-3)}|\xi|)$ has a cusp at $\xi=0$,
with an arbitrary height $A>0$, 
while its phase 
$\arg(u)= \phi -\epsilon \sqrt{c/(\sigma^2-3)}\xi + \epsilon 2(\sigma^2+1)\sqrt{c/(\sigma^2-3)}^3 t$ (modulo $2\pi$)
exhibits a jump discontinuity at $t\neq 0$ yet is continuous at $t=0$. 
Thus, this solution is a type of singular peakon. 

\begin{thm}\label{linphasekinks}
For a complex mKdV equation (\ref{mkdveqn}), 
non-singular kink solutions 
having a non-zero linear phase of the form (\ref{LINEARPHASE}) 
are admitted only in the case
$\Im \alpha =0$, $\beta=0$:
\begin{equation}\label{KinkSolutions1}
u(t,x)=
\sqrt{\frac{3(c+3k^2)}{\alpha}}\exp(i\phi)\exp(ik(x-(3c+8k^2)t))\tanh(\sqrt{-(c+3k^2)/2}(x-ct))
\end{equation}
where $c<-3k^2$, $\alpha<0$, and the parameters are given by 
$-\infty<k<\infty$, $0\leq\phi<2\pi$. 
\end{thm}

Solution (\ref{KinkSolutions1}) generalizes the kink solution given by $k=0$ for the integrable case 
$\alpha=\bar{\alpha}, \beta=0$ in the class of complex mKdV equations (\ref{mkdveqn}).

Further discussion of the solutions (\ref{SWsolutions1}), (\ref{SWsolutions3}), (\ref{KinkSolutions1})
will be given in section \ref{solnfeatures}.

\subsection{Computational Remarks}\label{Summarize2}

The integrability analysis of the overdetermined ODE system
(\ref{eqns1})--(\ref{eqns2})
is non-trivial due to the nonlinearity of the two ODEs
and the number of parameters in each ODE.

The analysis begins by using differentiation combined with cross-multiplication 
to eliminate $f'''$, $f''$, $f'$ until a purely algebraic equation 
in terms of $f$ is obtained. 
During elimination steps, a case distinction arises 
whenever the coefficient of a leading derivative can possibly vanish. 
These coefficients involve the parameters 
$\alpha_1$, $\alpha_2$, $\beta_1$, $\beta_2$, $k$, $w$ 
and sometimes also the functions $f$ and $f'$, 
and so each case distinction increases the size of the system. 
Because the original ODEs (\ref{eqns1}) and (\ref{eqns2}) are autonomous,
the variable $\xi$ does not appear explicitly 
in the coefficients or the equations. 
Consequently, 
any resulting equation that is algebraic in $f$ 
can be split with respect to all powers of $f$, 
which gives algebraic conditions just 
on $\alpha_1$, $\alpha_2$, $\beta_1$, $\beta_2$, $k$, $w$.
These conditions are first reduced with each other 
and then used to reduce the remaining equations in the system. 

Because of the constantly increasing number of options 
of how to proceed in the computation 
(i.e.\ the number of equations grows due to the repeated case distinctions, 
offering an increasing number of possible reductions to do, 
which in turn generate new equations), 
it is difficult for a computer program to perform the computations automatically. 
Computer algebra programs designed for 
differential Gr\"{o}bner basis computations 
(i.e.\ cross-differentiations and cross-reductions of differential equations
based on an ordering of derivatives 
which guarantees the finiteness of the algorithm for linear problems) 
are typically designed and tuned to do a kind of ``breadth-first search'' 
that aims to prevent or delay the generation of very large equations. 
But the situation for the present computation is different in that 
even a large algebraic equation involving $f$ is very useful 
as it can be directly split (because the system is autonomous), 
yielding useful shorter algebraic conditions that involve only 
$\alpha_1$, $\alpha_2$, $\beta_1$, $\beta_2$, $k$, $w$.
Therefore, it is much more productive to employ a ``depth-first search'' 
aiming at reducing the differential order of one equation as quickly 
as possible in the computation. 

For these reasons, the computer algebra package {\sc Crack} \cite{crack} has been used. 
The numerous techniques (modules) in this package 
for solving overdetermined systems of differential/algebraic equations
include automatic and interactive eliminations, substitutions, 
length-shortening of equations, and factorizations, among others. 
In particular, the ability to use the program interactively is essential
and gave the complete solution of the integrability analysis. 
A streamlined version of this interactive run 
(when executed in an automatic mode) 
used about 300 interactive steps taking 4 seconds in total on a 3GHz PC.
In other words, 
the computation is short for a computer program if it is guided properly, 
but it is long enough that a hand computation would be undesirable. 
(By comparison, the Maple program {\sc RiffSimp} running on a workstation 
for one day was unable to complete the full computation.)

As a result of tackling this application,
the package {\sc Crack} has been enhanced to perform an automatic splitting 
on equations that contain powers of any function occurring only polynomially 
and having a non-vanishing derivative with respect to at least one variable 
such that this variable neither appears explicitly in the equation
nor occurs as a variable of another function in the equation.

\section{Conservation Laws}\label{conlaws}
In the class of complex mKdV equations (\ref{mkdveqn}), 
all conservation laws (\ref{conslaw})
can be expressed in the integrated form 
\begin{equation}\label{integralC}
\frac{d}{dt}C = - X \big|_{-\infty}^{+\infty}
\end{equation}
in terms of corresponding conserved quantities $C=\int_{-\infty}^\infty T\,dx$, 
where the conserved density $T$ and flux $X$ are functions of 
$t$, $x$, $u$, $\bar{u}$, and $x$-derivatives of $u$, $\bar{u}$ 
(up to some finite order) 
after eliminating $t$-derivatives of $u$, $\bar{u}$ 
through equation (\ref{mkdveqn}). 
Two conservation laws are equivalent if their conserved densities $T$ 
differ by a total $x$-derivative $D_x \Upsilon$ 
whenever $u$ satisfies equation (\ref{mkdveqn}),
so that the same conserved quantity $C$ is obtained up to boundary terms 
$\Upsilon \big|_{-\infty}^{+\infty}$,
where $\Upsilon$ is any function of $t$, $x$, $u$, $\bar{u}$, 
and $x$-derivatives of $u$, $\bar{u}$. 
Correspondingly, the fluxes $X$ of two equivalent conservation laws 
differ by a total $t$-derivative $-D_t \Upsilon$
whenever $u$ satisfies equation (\ref{mkdveqn}). 
Since any conserved quantity can be decomposed into separately conserved
real and imaginary parts, there is no loss of generality in considering
only real-valued conserved densities and fluxes. 
The set of all such conservation laws up to equivalence 
for a given complex mKdV equation (\ref{mkdveqn}) forms a real vector space,
on which there is a natural symmetry group action generated by 
scalings (\ref{scaling}), time translations (\ref{timetrans}), 
space translations (\ref{spacetrans}), and phase rotations (\ref{phaserot}).

The ordinary mKdV equation has well-known conservation laws 
\cite{MiuGarKru} corresponding to conserved quantities 
for mass $\int_{-\infty}^\infty u\,dx$, 
momentum $\int_{-\infty}^\infty u^2\,dx$, 
energy $\int_{-\infty}^\infty (3u_x^2-u^4)\,dx$, 
and Galilean energy $\int_{-\infty}^\infty (3tu_x^2-tu^4+xu^2)\,dx$.
We will be interested, first, 
in finding all counterparts of these conserved quantities 
for complex mKdV equations (\ref{mkdveqn}) 
through a classification of all real-valued conserved densities of the form 
\begin{equation}\label{Cdensity}
T(t,x,u,\bar{u},u_x,\bar{u}_x)
\end{equation} 
modulo total $x$-derivatives.

Each conservation law (\ref{conslaw}) of a complex mKdV equation 
in which the conserved density $T$ and flux $X$ are real-valued expressions
can be expressed in an equivalent characteristic form \cite{2ndbook} 
\begin{equation}\label{ConDef}
D_tT+D_x (X-\Gamma) = 2\Re((u_t+\alpha \bar{u}uu_x+\beta u^2\bar{u}_x+u_{xxx})\bar{Q})
\end{equation}
given by a complex multiplier $Q=Q_1+iQ_2$, 
where the additional flux terms $\Gamma$ are linear in 
$u_t$, $\bar{u}_t$ and $x$-derivatives of $u_t$, $\bar{u}_t$ 
(up to a finite order), 
with $u$ now taken to be an arbitrary function of $t$, $x$. 
The characteristic form (\ref{ConDef}) is useful 
because it leads to a linear determining system for all conservation laws 
(up to equivalence) as follows. 

First, 
conserved densities $T$ modulo total $x$-derivatives have a one-to-one
correspondence to multipliers $Q$ through the variational expressions 
\begin{equation}\label{Qrelations}
Q = \frac{\delta T}{\delta \bar{u}}, \quad  \bar{Q} = \frac{\delta T}{\delta u}
\end{equation}
where $\delta/{\delta u}$ and $\delta/{\delta \bar{u}}$ denote variational derivatives
(Euler operators) with respect to $u$ and $\bar{u}$. 
As a result, $Q$ is a function of $t$, $x$, $u$, $\bar{u}$, 
and $x$-derivatives of $u$, $\bar{u}$ 
(up to twice the order of $T$ modulo $D_x \Upsilon$). 
Second, 
all multipliers $Q$ are determined by a linear system \cite{Olvbook,2ndbook} 
\begin{equation}\label{Qlsystem}
 \frac{\delta}{\delta u}\Re((u_t+\alpha \bar{u}uu_x+\beta u^2\bar{u}_x+u_{xxx})\bar{Q})
 =\frac{\delta}{\delta \bar{u}} \Re((u_t+\alpha \bar{u}uu_x+\beta u^2\bar{u}_x+u_{xxx})\bar{Q})=0
\end{equation}
in which $u$ is an arbitrary function of $t$, $x$. 
Since $Q$ has no dependence on $u_t$, $\bar{u}_t$, and their $x$-derivatives,
this system (\ref{Qlsystem}) has a natural splitting 
with respect to these variables.
The resulting split system consists of \cite{AncBlu2002a,AncBlu2002b} 
the adjoint of the determining equation for infinitesimal symmetries, 
augmented by the Helmholtz integrability equations for variational (Euler-Lagrange) expressions. 
Consequently, 
the multipliers $Q$ for a given complex mKdV equation (\ref{mkdveqn}) 
can be characterized as adjoint-symmetries that have a variational form. 
Third, 
each multiplier $Q$ determines a corresponding conserved density $T$ 
modulo $D_x \Upsilon$ through the variational relation (\ref{Qrelations}).
In particular, this relation can be explicitly inverted to obtain $T$ in terms of $Q$ 
by means of a homotopy integration formula \cite{2ndbook}
\begin{equation}\label{IFormula}
T=2\Re(\bar{u} \int_0^1  Q[\lambda u]\,d\lambda)+D_x \Upsilon
\end{equation}
where $Q[\lambda u]$ denotes the function $Q$ evaluated with 
$u$, $\bar{u}$, and all $x$-derivatives of $u$, $\bar{u}$ 
replaced by $\lambda u$, $\lambda \bar{u}$, and their $x$-derivatives. 

Now, for classifying conserved densities (\ref{Cdensity}), 
we will need to find all multipliers of the form 
\begin{equation}\label{Qrelations2}
Q(t,x,u,\bar{u},u_x,\bar{u}_x,u_{xx},\bar{u}_{xx}) . 
\end{equation}
The linear determining system for such multipliers is then given by 
the determining equation for adjoint-symmetries 
\begin{equation}\label{Ldsystem}
D_t Q +(2\beta-\alpha)uu_x \bar{Q}+(\bar{\alpha}-2\bar{\beta})\bar{u}u_xQ+\bar{\alpha}\bar{u}uD_xQ
+\beta u^2D_x \bar{Q}+D_x^3Q= 0
\end{equation}
and its complex conjugate, augmented by the Helmholtz integrability equations
\begin{subequations}\label{Hiequations}
\begin{align}
&
D_x \Re \frac{\partial Q}{\partial \bar{u}_{xx}} = \Re\frac{\partial Q}{\partial \bar{u}_{x}}, \quad
D_x \Im\frac{\partial Q}{\partial \bar{u}_{xx}} = \Im\frac{\partial Q}{\partial \bar{u}_{x}} ,
\\
&
\Im \frac{\partial Q}{\partial u_{xx}}=0, \quad D_x \Re \frac{\partial Q}{\partial u_{xx}} =\Re \frac{\partial Q}{\partial u_{x}}, \quad D_x \Im \frac{\partial Q}{\partial u_{x}}=2\Im \frac{\partial Q}{\partial u} . 
\end{align}
\end{subequations}
In equation (\ref{Ldsystem}), 
the variables $u_t$, $\bar{u}_t$, and their $x$-derivatives 
are eliminated via equation (\ref{mkdveqn}) and its $x$-derivatives 
(and their complex conjugates).
Thus, the system (\ref{Ldsystem})--(\ref{Hiequations}) 
will involve the variables $t$, $x$, $u$, $\bar{u}$,
and $x$-derivatives of $u$, $\bar{u}$ up to 5th order,
with the unknowns consisting of $Q$, $\alpha$, $\beta$. 
Since $Q$ is at most 2nd order, 
this system will split with respect to 
$u_{xxx}$, $\bar{u}_{xxx}$, $u_{xxxx}$, $\bar{u}_{xxxx}$, $u_{xxxxx}$, $\bar{u}_{xxxxx}$ 
into an overdetermined system of equations. 
We use computer algebra both to carry out the splitting
and to solve the resulting overdetermined system, 
as summarized in section \ref{CRemark}, 
which leads to the following classification result.

\begin{prop}\label{2ndQsolns}
The determining system (\ref{Ldsystem})--(\ref{Hiequations}) for 2nd order multipliers (\ref{Qrelations2}) has eight solutions:
\begin{subequations}\label{2ndQsol1}
\begin{align}
&Q=1\\
\intertext{with}
&\alpha_1=2\beta_1, \alpha_2=2\beta_2 ;
\end{align}
\end{subequations}
\begin{subequations}\label{2ndQsol2}
\begin{align}
&Q=i\\
\intertext{with}
&\alpha_1=2\beta_1, \alpha_2=2\beta_2 ;
\end{align}
\end{subequations}
\begin{subequations}\label{2ndQsol3}
\begin{align}
&Q=u\\
\intertext{with}
&\alpha_2=\beta_2;
\end{align}
\end{subequations}
\begin{subequations}\label{2ndQsol4}
\begin{align}
&Q=i\bar{u}\\
\intertext{with}
&\alpha_1=3\beta_1, \alpha_2=3\beta_2;
\end{align}
\end{subequations}
\begin{subequations}\label{2ndQsol5}
\begin{align}
&Q=\bar{u}\\
\intertext{with}
&\alpha_1=3\beta_1, \alpha_2=3\beta_2;
\end{align}
\end{subequations}
\begin{subequations}\label{2ndQsol6}
\begin{align}
&Q=iu_x\\
\intertext{with}
&\alpha_2=\beta_1=\beta_2 =0;
\end{align}
\end{subequations}
\begin{subequations}\label{2ndQsol7}
\begin{align}
&Q=3u_{xx}+(\alpha_1+\beta_1)u^2\bar{u}\\
\intertext{with}
&\alpha_2=\beta_2=0;
\end{align}
\end{subequations}
\begin{subequations}\label{2ndQsol8}
\begin{align}
&Q=3tu_{xx}+(\alpha_1+\beta_1)tu^2\bar{u}-xu\\
\intertext{with}
&\alpha_2=\beta_2=0.
\end{align}
\end{subequations}
\end{prop}

The conserved densities (\ref{Cdensity}) 
corresponding to the multipliers (\ref{2ndQsol1})--(\ref{2ndQsol8}) 
can be obtained from the integration formula (\ref{IFormula}).
One drawback of this formula is that 
if a multiplier $Q$ contains $x$-derivatives of $u$ or $\bar{u}$
then the resulting conserved density does not necessarily have the lowest possible order.
In particular, if $Q$ contains $u_{xx}$ or $\bar{u}_{xx}$, 
then $T$ will have terms containing $u_{xx}$ and $\bar{u}_{xx}$, 
which can be canceled by subtraction of suitable $D_x \Upsilon$ terms.
Through the Helmholtz integrability equations (\ref{Hiequations}), 
we can show that these terms are given by the integral
\begin{equation}
\Upsilon = 
2\Re\int\big(
\bar{u} \int_0^1 \frac{\partial Q[\lambda u]}{\partial u_{xx}}\frac{d\lambda}{\lambda}
+u\int_0^1 \frac{\partial \bar{Q}[\lambda u]}{\partial u_{xx}}\frac{d\lambda}{\lambda}
\big)\,du_x
\end{equation}
Hence we obtain an improved homotopy integration formula
\begin{equation}\label{HIFormula}
 T= 
2\Re\int_0^1 \big(\lambda \bar{u}Q_0(t,x,\lambda u,\lambda \bar{u})
-W(t,x,\lambda u,\lambda \bar{u},\lambda u_x,\lambda \bar{u}_x)\big)
\frac{d\lambda}{\lambda}
\end{equation}
where 
\begin{align}
\begin{split}
W= 
u_x \int_0^1\big( & 
\lambda \bar{u}_x \frac{\partial Q}{\partial u_{xx}}(t,x,u,\bar{u},\lambda u_x,\lambda \bar{u}_x,0,0)
+\lambda u_x \frac{\partial \bar{Q}}{\partial u_{xx}}(t,x,u,\bar{u},\lambda u_x,\lambda \bar{u}_x,0,0)\\
&\quad 
-i\bar{u}\Im \frac{\partial Q_1}{\partial u_{x}}(t,x,u,\bar{u},\lambda u_x,\lambda \bar{u}_x)
\big)\, d\lambda
\end{split}
\end{align}
in terms of 
$Q_1 = Q(t,x,u,\bar{u},u_x,\bar{u}_x,0,0)$ 
and $Q_0 = Q_1(t,x,u,\bar{u},0,0)$. 
Evaluation of this formula (\ref{HIFormula}) for the multipliers (\ref{2ndQsol1})--(\ref{2ndQsol8}) 
leads to the following main result.  

\begin{thm}\label{1storderT}
For a complex mKdV equation (\ref{mkdveqn}), 
conserved densities of the form (\ref{Cdensity}) are admitted only in the following cases:
\newline
{\rm (i)} 
$\Im \alpha = \Im \beta$
\begin{equation}\label{T1}
T=u\bar{u} .
\end{equation}
\newline
{\rm (ii)} 
$\Im \alpha =0, \Im \beta =0$
\begin{align}
&T=-3u_x\bar{u}_x+\tfrac{1}{2}(\alpha+\beta)u^2\bar{u}^2 ;
\label{T2}\\
&T=-3tu_x\bar{u}_x+\tfrac{1}{2}(\alpha+\beta) tu^2\bar{u}^2-xu\bar{u} .
\label{T3}
\end{align}
\newline
{\rm (iii)} 
$\alpha = 2\beta$
\begin{align}
&T=u+\bar{u} ;
\label{T4}\\
&T=i(\bar{u}-u) .
\label{T5}
\end{align}
\newline
{\rm (iv)} 
$\alpha = 3\beta$
\begin{align}
&T=\tfrac{1}{2}(u^2+\bar{u}^2) ;
\label{T6}\\
&T=i\tfrac{1}{2}(\bar{u}^2-u^2) .
\label{T7}
\end{align}
\newline
{\rm (v)} 
$\Im \alpha = 0, \beta =0$
\begin{equation}\label{T8}
T=i(u_{x}\bar{u}-\bar{u}_xu) .
\end{equation}
\end{thm}

As indicated by their form, 
the conserved densities (\ref{T1}), (\ref{T2}) and (\ref{T3}) 
are phase-invariant counterparts of the respective conserved densities 
for momentum, energy and Galilean energy listed earlier 
for the ordinary (real) mKdV equation. 
The form of the other phase-invariant conserved density (\ref{T8}) 
suggests that it has the meaning of an angular (phase) twist,
since it vanishes when the phase of $u$ is constant. 
The remaining conserved densities (\ref{T4}), (\ref{T5}), (\ref{T6}), (\ref{T7})
are not phase-invariant. 
In particular, conserved densities (\ref{T4}) and (\ref{T5})
are related by a phase rotation 
$u\rightarrow \exp(i\pi/2)u$, 
while conserved densities (\ref{T6}) and (\ref{T7}) are similarly related by 
$u\rightarrow \exp(i\pi/4)u$.
This suggests that the following complex linear combinations of
these densities can be viewed as a phase-covariant mass density
\begin{equation}\label{CDens1}
T=u
\end{equation}
 and a phase-covariant momentum density
\begin{equation}\label{CDens2}
T=u^2
\end{equation}
where the complex conserved densities 
(\ref{CDens1}) and (\ref{CDens2}) are homogeneous with respect to 
phase rotations (\ref{phaserot}) on $u$.  
Further discussion of the conserved quantities defined by 
all of the phase-invariant densities 
(\ref{T1}), (\ref{T2}), (\ref{T3}), (\ref{T8})
and phase-covariant densities (\ref{CDens1}), (\ref{CDens2}) 
will be given in section \ref{solnfeatures}.

Finally, 
we will classify all real-valued higher derivative conserved densities 
having the lowest-order form 
\begin{equation}\label{2ndCdens}
T(t,x,u,\bar{u},u_x,\bar{u}_x,u_{xx},\bar{u}_{xx})
\end{equation}
with an essential dependence on $u_{xx}$ and $\bar{u}_{xx}$, 
modulo total $x$-derivatives.
The existence of such conserved densities will be important for detecting 
integrable cases in the class of complex mKdV equations (\ref{mkdveqn}). 
For this classification, we need to find all multipliers of the form
\begin{equation}\label{4thQ}
Q(t,x,u,\bar{u},u_x,\bar{u}_x,u_{xx},\bar{u}_{xx},u_{xxx},\bar{u}_{xxx},u_{xxxx},\bar{u}_{xxxx})
\end{equation}
with an essential dependence on $u_{xxxx}$, $\bar{u}_{xxxx}$, or $u_{xxx}$, $\bar{u}_{xxx}$, 
satisfying the linear determining system (\ref{Qlsystem}). 
Again, this system splits into the adjoint-symmetry equation (\ref{Ldsystem}) 
and its complex conjugate, 
augmented by the Helmholtz integrability equations which now are given by
\begin{subequations}\label{HIequations1}
\begin{align}
&
2D_x \Re\frac{\partial Q}{\partial \bar{u}_{xxxx}} =\Re\frac{\partial Q}{\partial \bar{u}_{xxx}},
\\
&
2D_x \Im \frac{\partial Q}{\partial \bar{u}_{xxxx}} =\Im\frac{\partial Q}{\partial \bar{u}_{xxx}},
\\
&
D_x^2 \Re \frac{\partial Q}{\partial \bar{u}_{xxx}}-2D_x \Re\frac{\partial Q}{\partial \bar{u}_{xx}} = -2\Re\frac{\partial Q}{\partial \bar{u}_{x}},
\\
&
D_x^2 \Im \frac{\partial Q}{\partial \bar{u}_{xxx}}-2D_x \Im\frac{\partial Q}{\partial \bar{u}_{xx}} =-2\Im\frac{\partial Q}{\partial \bar{u}_{x}} ,
\\
&
\Im \frac{\partial Q}{\partial u_{xxxx}} = 0,
\\
&
2D_x \Re \frac{\partial Q}{\partial u_{xxxx}}= \Re \frac{\partial Q}{\partial \bar{u}_{xxx}} , 
\\
&
3D_x \Im \frac{\partial Q}{\partial u_{xxx}}= 2\Im \frac{\partial Q}{\partial u_{xx}}, 
\\ 
&
D_x^2 \Re \frac{\partial Q}{\partial u_{xxx}}-2D_x \Re \frac{\partial Q}{\partial u_{xx}} =-2\Re \frac{\partial Q}{\partial u_{x}},
\\
&
D_x^2 \Im \frac{\partial Q}{\partial u_{xx}}-3D_x \Im \frac{\partial Q}{\partial u_{x}}=-6\Im \frac{\partial Q}{\partial u} .
\end{align}
\end{subequations}
The determining system (\ref{Ldsystem}) and (\ref{HIequations1}) 
involves the variables $t$, $x$, $u$, $\bar{u}$, and $x$-derivatives of $u$, $\bar{u}$ up to 7th order. 
Since $Q$ is at most 4th order, 
this system will split into a large overdetermined system of equations, 
which we derive and solve by using computer algebra, 
as summarized in section \ref{CRemark}. 
We thereby obtain the following classification result.

\begin{prop}\label{4thQsolns}
The determining system (\ref{Ldsystem}) and (\ref{HIequations1}) 
for 3rd and 4th order multipliers (\ref{4thQ}) has three solutions:
\begin{subequations}\label{4thQsol1}
\begin{align}
&Q=i(u_{xxx}+\alpha_1 u\bar{u}u_x)\\
\intertext{with}
&\alpha_2=\beta_1=\beta_2=0 ;
\end{align}
\end{subequations}
\begin{subequations}\label{4thQsol2}
\begin{align}
&Q=6u_{xxxx}+2\alpha_1 u^2\bar{u}_{xx}+8\alpha_1 u\bar{u}u_{xx}+6\alpha_1 \bar{u}u_x^2+4\alpha_1 uu_x\bar{u}_x+\alpha_1^2 u^3\bar{u}^2 \\
\intertext{with}
&\alpha_2=\beta_1=\beta_2=0 ;
\end{align}
\end{subequations}
\begin{subequations}\label{4thQsol3}
\begin{align}
&Q=3u_{xxxx}+6\beta_1 u^2\bar{u}_{xx}+14\beta_1 u\bar{u}u_{xx}+8\beta_1 \bar{u}u_x^2
+12\beta_1 u\bar{u}_xu_x+8\beta_1^2 u^3\bar{u}^2 \\
\intertext{with}
&\alpha_1=3\beta_1, \quad \alpha_2=\beta_2=0 .
\end{align}
\end{subequations}
\end{prop}

To obtain the 2nd order conserved densities (\ref{2ndCdens}) 
determined by these multipliers (\ref{4thQsol1})--(\ref{4thQsol3}), 
we evaluate the integration formula (\ref{IFormula}) 
and cancel all higher-order terms (containing 
$u_{xxx}$, $\bar{u}_{xxx}$, $u_{xxxx}$, $\bar{u}_{xxxx}$) 
by subtracting suitable total $x$-derivatives. 
This yields the following classification result.

\begin{thm}\label{TTS}
For a complex mKdV equation (\ref{mkdveqn}), 
conserved densities of the form (\ref{2ndCdens})
are admitted only in the following cases:
\newline
{\rm (i)} 
$\Im \alpha = 0, \beta=0$
\begin{align}
T=&
i\tfrac{1}{2}(\bar{u}_{xx}u_{x}-u_{xx}\bar{u}_{x})+i6\alpha (u\bar{u}^2u_x-u^2\bar{u}\bar{u}_x) ;
\label{4thT1}\\
\begin{split}\label{4thT2}
T=&
6u_{xx}\bar{u}_{xx}-\tfrac{1}{4}\alpha (u+\bar{u})(u-\bar{u})^2(u_{xx}+\bar{u}_{xx})
+\tfrac{1}{4}\alpha (2u\bar{u}-3u^2-3\bar{u}^2)(u_x^2+\bar{u}_x^2)
\\
&\quad -\tfrac{1}{2}\alpha (u^2+14u\bar{u}+\bar{u}^2)u_x\bar{u}_x+\tfrac{1}{3}\alpha ^2u^3\bar{u}^3 .
\end{split}
\end{align}
\newline
{\rm (ii)} 
$\alpha=3\beta, \Im \alpha =\Im \beta=0$
\begin{align}
\begin{split}\label{4thT3}
T=&
3u_{xx}\bar{u}_{xx}-\tfrac{3}{4}\beta (u+\bar{u})(u-\bar{u})^2(u_{xx}+\bar{u}_{xx})
-\tfrac{1}{12}\beta (27u^2-18u\bar{u}+27\bar{u}^2)(u_x^2+\bar{u}_x^2)
\\
&\quad
-\tfrac{1}{2}\beta (22u\bar{u}+3u^2+3\bar{u}^2) u_x\bar{u}_x+\tfrac{4}{3}\beta^2 u^3\bar{u}^3 .
\end{split}
\end{align}
\end{thm}

The two cases here in which complex mKdV equations admit 
these higher-derivative conserved densities
are exactly the well-known integrable cases 
$\alpha/\beta =0$ and $\alpha/\beta=3$ with real coefficients,
$\alpha =\bar{\alpha}$, $\beta =\bar{\beta}$. 
This result indicates that any new integrable cases for
complex mKdV equations having complex coefficients
will involve a recursion operator of order greater than two.

\subsection{Computational Remarks}\label{CRemark}
For computational purposes, 
it is convenient to replace the complex variables and unknowns by 
their real and imaginary components, namely, 
$u=u_1+iu_2, u_x=u_{1x}+iu_{2x}$, 
$Q=Q_1+iQ_2,\alpha=\alpha_1+i\alpha_2,\beta=\beta_1+i\beta_2$. 
In component form, the characteristic form of a conservation law (\ref{ConDef}) is given by
\begin{align}
\begin{split}
D_tT+D_xX =&
(u_{1t}+((\alpha_1+\beta_1)u_1^2-2\beta_2u_1u_2+(\alpha_1-\beta_1)u_2^2)u_{1x}
-((\alpha_2-\beta_2)u_1^2-2\beta_1u_1u_2
\\&\quad 
+(\alpha_2+\beta_2)u_2^2)u_{2x}+u_{1xxx})Q_1
+(u_{2t}+((\alpha_2+\beta_2)u_1^2 +2\beta_1u_1u_2
\\&\quad 
+(\alpha_2-\beta_2)u_2^2)u_{1x} +((\alpha_1-\beta_1)u_1^2+2\beta_2u_1u_2
+(\alpha_1+\beta_1)u_2^2)u_{2x}+u_{2xxx})Q_2
\end{split}
\end{align}
in terms of a multiplier pair $(Q_1, Q_2)$, 
while the variational expressions (\ref{Qrelations})
relating such multipliers to conserved densities take the form 
\begin{equation}
Q_1 = \frac{\delta T}{\delta u_1},\quad
Q_2 = \frac{\delta T}{\delta u_2} . 
\end{equation}

For 2nd order multipliers
\begin{equation}
Q_1(t,x,u_1,u_2,u_{1x},u_{2x},u_{1xx},u_{2xx}), 
\quad
Q_2(t,x,u_1,u_2,u_{1x},u_{2x},u_{1xx},u_{2xx})
\end{equation}
the determining system (\ref{Ldsystem})--(\ref{Hiequations}) 
splits into an overdetermined system of 53 real equations
for the 6 unknowns $\alpha_1,\alpha_2,\beta_1,\beta_2, Q_1, Q_2$.
The main complication in solving the system is that 
it contains terms in which $Q_1$ and $Q_2$ appear 
with $\alpha_1$, $\alpha_2$, $\beta_1$, $\beta_2$ as coefficients, 
so that the system as a whole is nonlinear in the joint unknowns. 
To formulate and solve this system, 
we use the program {\sc Conlaw} \cite{conlaw}, 
which in turn calls the package {\sc Crack} \cite{crack} 
for solving overdetermined systems of differential-algebraic equations. 

{\sc Crack} uses a wide repertoire of techniques (modules), including 
eliminations, substitutions, integrations, direct and indirect separations, 
length-shortening of equations, and factorizations.
In the present computation, 
the typical techniques needed are integration, substitution, and separation. 
Integration of equations is not hindered by the nonlinearity of the system,
because $\alpha_1$, $\alpha_2$, $\beta_1$, $\beta_2$ are just constants 
while $Q_1$ and $Q_2$ appear only linearly. 
In contrast, indirect separation of equations is sensitive to the nonlinearity.
An equation is indirectly separable when no unknown in the equation 
depends on all independent variables but each independent variable occurs 
in at least one unknown in the equation, 
i.e.\ direct splitting is therefore not possible. 
Such equations can be separated by a sequence of divisions and differentiations,
where each division requires a case distinction 
depending on whether the divisor is zero or not. 
For nonlinear equations, 
a simple alternative which has been implemented in {\sc Crack} 
is to differentiate the equation repeatedly and perform case generating substitutions and case generating Gr\"{o}bner steps. 
The entire computation to obtain all solutions takes about 1600 steps
and is essentially interactive. 

For 4th order multipliers
\begin{align}
\begin{split}
&Q_1(t,x,u_1,u_2,u_{1x},u_{2x},u_{1xx},u_{2xx},u_{1xxx},u_{2xxx},u_{1xxxx},u_{2xxxx}),\\ 
&Q_2(t,x,u_1,u_2,u_{1x},u_{2x},u_{1xx},u_{2xx},u_{1xxx},u_{2xxx},u_{1xxxx},u_{2xxxx})
\end{split}
\end{align}
the determining system (\ref{Ldsystem}) and (\ref{HIequations1})
splits into a larger overdetermined system of 212 real equations.
The resulting system contains, 
on the one hand, many equations that are 
linear, short (less than 100 terms), and easy to integrate, 
and on the other hand, 
equations that are longer (up to almost 38000 terms) and nonlinear. 
Because of this complexity, 
the whole computation to solve this system takes several thousand steps 
consisting of integrations, substitutions, separations, decouplings, and factorizations,
while the steps that involve substitutions and separations turn out to be very time consuming when performed on the long equations. 
This prompted the following two enhancements to {\sc Crack} 
for speeding up the computation. 

First, 
a more specialized separation module has been added as a default module.
This module performs only direct separations 
(which are the most frequently needed splittings), 
whereas the general module also performs indirect separations as well as splittings 
where unknowns occur in exponents of independent variables. 
Splittings with respect to exponents are relatively slow, 
because this involves converting expressions internally between two different forms 
(prefix form and standard quotient form). 
Since none of the unknowns in the present computation occur in exponents,
a significant speed-up is achieved by avoiding the use of the general separation module. 

Second, 
a new module that partially solves subsystems has been developed, 
producing another major speed-up. 
Previously, each substitution resulting from each integrated equation 
had to be performed in all equations in the system at once, 
which gets very time consuming when the system contains long equations. 
The new module allows short, linear equations to be integrated and/or split, 
followed by substitution into other short equations, 
so that after several such iterations the equations in the subsystem
become much shorter, have low order, and are practically all nonlinear in the unknowns.
At that stage the preliminary solution is then
substituted in the whole system including long equations.

Two other features of {\sc Crack}  greatly aid the present computation. 
{\sc Crack} is able to collect inequalities 
(actively generated from conditions on the unknowns 
in combination with algebraic equations that arise during the computation) 
and use them to select steps that will minimize the number of case distinctions and keep the 
depth of case nestings as low as possible. 
Finally, {\sc Crack} can be run interactively as well as automatically,	
which is especially useful in solving large nonlinear problems.

\section{Conserved quantities for solitary waves and kinks}\label{solnfeatures}

The conserved densities obtained in Theorem~\ref{1storderT}
for the class of complex mKdV equations (\ref{mkdveqn}) 
yield conservation laws with the following integrated form (\ref{integralC}), 
where
\begin{equation}\label{IDensity}
C=\int_{-\infty}^\infty T(t,x,u,\bar{u},u_x,\bar{u}_x)\, dx
\end{equation}
is the integrated density and 
\begin{equation}\label{CFlux}
X(t,x,u,\bar{u},u_x,\bar{u}_x,u_{xx},\bar{u}_{xx},u_{xxx},\bar{u}_{xxx})
\end{equation}
is the corresponding flux:
\newline
(i) $\Im \alpha = \Im \beta$ 
\begin{subequations}\label{mom}
\begin{align}
&C=\int_{-\infty}^\infty |u|^2\, dx 
=\mathcal{P} 
\qquad \qquad \text{(momentum)}\\
&{\rm X} =\tfrac{1}{2}(\alpha+\beta)u^2\bar{u}^2-u_x\bar{u}_x+u\bar{u}_{xx}+\bar{u}u_{xx} ;
\end{align}
\end{subequations}
\newline
(ii) $\Im \alpha = 0,  \Im \beta = 0$
\begin{subequations}\label{ener}
\begin{align}
&C=\int_{-\infty}^\infty -3|u_x|^2+\tfrac{1}{2}(\alpha+\beta)|u|^4 \,dx 
=2\mathcal{E} 
\qquad \qquad \text{(energy)}\\
\begin{split}
&{\rm X} =3(u_{xx}\bar{u}_{xx}-\bar{u}_xu_{xxx}-u_x\bar{u}_{xxx})+(\alpha+\beta)(u\bar{u}^2u_{xx}+u^2\bar{u}\bar{u}_{xx})-(5\alpha+2\beta)u\bar{u}u_x\bar{u}_x\\
&\qquad -\tfrac{1}{2}(\alpha+2\beta)(\bar{u}^2u_x^2+u^2\bar{u}_x^2)+\tfrac{1}{3}(\alpha+\beta)^2u^3\bar{u}^3 ;
\end{split}
\end{align}
\end{subequations}
\begin{subequations}\label{Galener}
\begin{align}
&C=\int_{-\infty}^\infty -3t|u_x|^2+\tfrac{1}{2}(\alpha+\beta)t|u|^4-x|u|^2\, dx 
=2\mathcal{G} 
\qquad \qquad \text{(Galilean energy)}\\
\begin{split}
&{\rm X} = u\bar{u}_x+\bar{u}u_x+x(u_x\bar{u}_x -\bar{u}u_{xx}-u\bar{u}_{xx})-\tfrac{1}{2}x(\alpha+\beta)u^2\bar{u}^2
-t(5\alpha+2\beta)u\bar{u}u_x\bar{u}_x\\
&\qquad -\tfrac{1}{2}t(\alpha+2\beta)(u_x^2\bar{u}^2+\bar{u}_x^2u^2)+3t(u_{xx}\bar{u}_{xx}-u_x\bar{u}_{xxx}-\bar{u}_xu_{xxx})\\
&\qquad +t(\alpha+\beta)(u_{xx}u\bar{u}^2+\bar{u}_{xx}\bar{u}u^2)+\tfrac{1}{3}t(\alpha+\beta)^2\bar{u}^3u^3 ;
\end{split}
\end{align}
\end{subequations}
\newline
(iii) $\alpha =2\beta $
\begin{subequations}\label{covmass}
\begin{align}
&C=\int_{-\infty}^\infty u\, dx 
=\tilde{\mathcal M}
\qquad \qquad \text{(phase-covariant mass)}\\
&{\rm X} = \beta u^2\bar{u}+u_{xx} ;
\end{align}
\end{subequations}
\newline
(iv) $\alpha =3\beta $
\begin{subequations}\label{covmom}
\begin{align}
&C=\int_{-\infty}^\infty u^2\, dx 
=\tilde{\mathcal P} 
\qquad \qquad \text{(phase-covariant momentum)}\\
&{\rm X} = 2\beta u^3\bar{u}-u_x^2+2uu_{xx} ;
\end{align}
\end{subequations}
\newline
(v) $\Im \alpha =0, \beta= 0$
\begin{subequations}\label{twist}
\begin{align}
&C=\int_{-\infty}^\infty -i|u|^2\arg(u)_x\, dx 
=\mathcal{W} 
\qquad \qquad \text{(angular twist)}\\
&{\rm X} =\alpha i(\bar{u}^2uu_x-\bar{u}u^2\bar{u}_x)+i\bar{u}u_{xxx}-iu\bar{u}_{xxx}+2i(u_x\bar{u}_{xx}-\bar{u}_xu_{xx}) .
\end{align}
\end{subequations}

For a given solution $u(t,x)$ of a complex mKdV equation, 
these integrated densities (\ref{mom})--(\ref{twist}) will define 
a finite conserved (i.e.\ time-independent) quantity (\ref{IDensity}) 
if and only if the densities $T(t,x,u,\bar{u},u_x,\bar{u}_x)$ themselves 
have sufficiently rapid decay as $|x| \rightarrow \infty$ 
and also the corresponding fluxes (\ref{CFlux}) 
vanish for $x \rightarrow \pm\infty$.

We will now determine which of the integrated densities yield 
a finite conserved quantity 
for the solitary wave solutions (\ref{solitary1})--(\ref{solitary4}), (\ref{SWsolutions1}) and the peakon solution (\ref{SWsolutions3}) 
obtained in Theorems~\ref{solitarywavesolns} and~\ref{linphasesolitarywaves} 
as well as for the kink solutions (\ref{kink1})--(\ref{kink2}), (\ref{KinkSolutions1}) 
obtained in Theorems~\ref{kinksolns} and~\ref{linphasekinks}.  
To proceed we first write down the asymptotic behaviour of each solution 
for large $\pm x$ such that $|\xi|\gg 1/\sqrt{|c|}$.
\newline\newline
\noindent
Solitary wave (\ref{solitary1}):
\begin{subequations}\label{SSOLUTIONS1}
\begin{align}
&u = e^{i(\phi+\theta)}\sqrt{\frac{c}{\alpha}}\tanh\Theta+O(\exp(-\sqrt{c} |x|))
\label{asymp1}\\
&\Im \alpha = 0, \beta = 0 . 
\end{align}
\end{subequations}
Solitary wave (\ref{solitary2}):
\begin{subequations}\label{SSOLUTIONS2}
\begin{align}
&u = e^{i(\phi+\theta)}\sqrt{\frac{c}{\alpha}}+O(1/|x|^2)
\label{asymp2}\\
&\Im \alpha = 0, \beta = 0 . 
\end{align}
\end{subequations}
Solitary wave (\ref{solitary3}):
\begin{subequations}\label{SSOLUTIONS3}
\begin{align}
&u = e^{i\phi}\sqrt{\frac{c}{\alpha+\beta}}\tanh\Theta+O(\exp(-\sqrt{c}|x|))
\label{asymp3}\\
&\Im (\alpha+\beta) = 0 . 
\end{align}
\end{subequations}
Solitary wave (\ref{solitary4}):
\begin{subequations}\label{SSOLUTIONS4}
\begin{align}
&u = e^{i\phi}\sqrt{\frac{c}{\alpha+\beta}}+O(1/|x|^2)
\label{asymp4}\\
&\Im (\alpha+\beta) = 0 . 
\end{align}
\end{subequations}
Solitary wave (\ref{SWsolutions1}):
\begin{subequations}\label{SSOLUTIONS5}
\begin{align}
&u =O(\exp(-\sqrt{c+3k^2}|x|))
\label{asympp7}\\
&\Im \alpha = 0, \beta = 0 . 
\end{align}
\end{subequations}
Peakon (\ref{SWsolutions3}):
\begin{subequations}\label{SSOLUTIONS6}
\begin{align}
&u =O(\exp(-\sqrt{c\sigma^2/(\sigma^2-3)}|x|))
\label{asympp8}\\
&|\alpha|=|\beta| 
\end{align}
where
\begin{equation} 
\sigma=\Re (\alpha-\beta)/\Im (\alpha+\beta)=-\Im (\alpha-\beta)/\Re (\alpha+\beta) . 
\end{equation}
\end{subequations}
Kink (\ref{kink1}):
\begin{subequations}\label{SSOLUTIONS7}
\begin{align}
&u = \pm e^{i\phi}\sqrt{\frac{3c}{\alpha+\beta}}+O(\exp(-\sqrt{-2c}|x|))
\label{asymp5}\\
&\Im (\alpha+\beta)= 0 . 
\end{align}
\end{subequations}
Kink (\ref{KinkSolutions1}):
\begin{subequations}\label{SSOLUTIONS8}
\begin{align}
&u =\pm e^{i\phi}\sqrt{\frac{3(c+3k^2)}{\alpha}}\exp(ik(x-(3c+8k^2)t))+O(\exp(-\sqrt{-2(c+3k^2)}|x|))
\label{asympp9}\\
&\Im \alpha=0, \beta= 0 . 
\end{align}
\end{subequations}
Kink (\ref{kink2}):
\begin{subequations}\label{SSOLUTIONS9}
\begin{align}
&u = e^{i\phi_\pm}\sqrt{{\frac{c}{\alpha}}(1+2\sech^2\Theta)}+O(\exp(-\sqrt{-2c}|x|))
\label{asymp6}\\
&\Im \alpha=0, \beta= 0 
\end{align}
where
\begin{equation}\label{asympphi} 
\tan(\phi_\pm - \phi)=\pm \tfrac{1}{\sqrt{3}}\sinh\Theta . 
\end{equation}
\end{subequations}

Based on this asymptotic behaviour, 
it is straightforward to derive the conditions on $\alpha$ and $\beta$ 
under which each integrated density (\ref{mom})--(\ref{twist}) is 
conserved and finite for each of the solutions. 
The conditions obtained for 
the solitary wave solutions (\ref{solitary1})--(\ref{solitary4}), (\ref{SWsolutions1}) and the peakon solution (\ref{SWsolutions3}) 
are listed in Table~\ref{coeffconds}
(where ``-'' means that an integrated density 
either is defined only for $\alpha=\beta=0$, 
or is infinite, or fails to be conserved). 
These conditions show that all six integrated densities 
are conserved and finite only for the solitary wave (\ref{solitary3}) 
in the special case $\Theta =0$ which reduces to 
the familiar $\sech$ solution (\ref{sechsoln}) 
having the asymptotic decay $u=O(\exp(-\sqrt{c}|x|))$ for large $\pm x$. 
For the peakon (\ref{SWsolutions3}), 
only the momentum (\ref{mom}) is conserved and finite
(while both the energy (\ref{ener}) and Galilean energy (\ref{Galener}) 
fail to be defined except in the trivial cases $\sigma=0$ or $1/\sigma =0$). 
In contrast, 
for the kink solutions (\ref{kink1}) and (\ref{kink2}), 
only the angular twist (\ref{twist}) is conserved and finite. 
The other kink solution (\ref{KinkSolutions1}) is exceptional 
as the linear phase term in its asymptotic behaviour (\ref{SSOLUTIONS8}) 
leads to the angular twist being neither finite nor conserved.
We now list the expressions for all of 
the conserved and finite integrated densities.
 
\begin{prop}\label{consquantities}
Solitary wave (\ref{sechsoln}) 
with speed $c>0$ and phase angle $0\leq\phi< 2\pi$ 
has the following conserved quantities:
\begin{align}
&\text{momentum}\qquad  
&{\mathcal P}(c) = \frac{12c^{1/2}}{\alpha+\beta} ,
\label{sechmom}\\
&\text{energy}\qquad   
&{\mathcal E}(c) = \frac{6c^{3/2}}{\alpha+\beta} ,
\label{sechener}\\
&\text{Galilean energy}\qquad
&{\mathcal G} = 0 ,
\label{sechGalener}\\
&\text{phase-covariant mass} \qquad  
&\tilde{\mathcal M}(\phi) = \frac{2\pi e^{i\phi}}{\sqrt{\alpha}} ,
\label{sechcovmass}\\
&\text{phase-covariant momentum} \qquad 
&\tilde{\mathcal P}(c,\phi)= \frac{9c^{1/2}e^{i2\phi}}{\alpha} ,
\label{sechcovmom}\\
&\text{angular twist} \qquad 
&\mathcal W=0 .
\label{sechtwist}
\end{align}
Solitary wave (\ref{SWsolutions1}) 
with speed $c>0$, phase angle $0\leq \phi<2\pi$, 
and frequency $-\infty< -k(3c+8k^2)<\infty$
has the following conserved quantities:
\begin{align}
&\text{momentum}\qquad  
&{\mathcal P}(c,k)= \frac{12(c+3k^2)^{1/2}}{\alpha} , 
\label{kmom}\\
&\text{energy}\qquad   
&{\mathcal E}(c,k) = \frac{6c(c+3k^2)^{1/2}}{\alpha} , 
\label{kener}\\
&\text{Galilean energy}\qquad
&{\mathcal G} = 0 , 
\label{kGalener}\\
&\text{angular twist} \qquad 
&\mathcal W=0 .
\label{ktwist}
\end{align}
\end{prop}

Interestingly, 
the momentum $\mathcal P$ and the energy $\mathcal E$ 
for these solitary waves (\ref{sechsoln}) and (\ref{SWsolutions1}) 
satisfy the relation 
\begin{equation}\label{MErelation}
{\mathcal E}= \tfrac{1}{2}c{\mathcal P} . 
\end{equation}
In addition, the Galilean energy is related to the momentum and energy by
\begin{equation}\label{GErelation}
2{\mathcal G}=2t{\mathcal E}- \chi(t){\mathcal P}=0
\end{equation}
where
\begin{equation}\label{CMomentum}
\chi(t)=\frac{1}{\mathcal P}\int_{-\infty}^\infty x|u|^2\, dx
\end{equation}
defines the center of momentum
(in analogy to the definition of center of mass). 
From relations (\ref{MErelation}) and (\ref{GErelation}), 
this time-dependent quantity (\ref{CMomentum}) is given by 
\begin{equation}\label{Krelation}
\chi(t)=ct . 
\end{equation}
Thus, both of the solitary waves (\ref{sechsoln}) and (\ref{SWsolutions1})
exhibit the same kinematic relations (\ref{MErelation}) and (\ref{Krelation}) 
obeyed by a free particle with speed $c$, 
whose position $x=ct$ coincides with the peak of $|u|$ 
as determined by $\xi=x-ct=0$.

\begin{prop}\label{constwist}
Solitary waves (\ref{solitary1})--(\ref{solitary4})
and kink (\ref{kink1}) 
have the conserved angular twist 
\begin{equation}\label{Angulartwist1}
\mathcal W=0 . 
\end{equation}
Kink (\ref{kink2}) has the conserved angular twist 
\begin{equation}\label{Angulartwist2}
\mathcal W(c,\Theta) = \frac{2\sqrt{3}c}{\alpha}\sinh\Theta\ \sech^2\Theta . 
\end{equation}
Solitary wave (\ref{SWsolutions1}) has the conserved angular twist
\begin{equation}\label{Angulartwist3}
\mathcal W(c,k) = \frac{-12k}{\alpha}\sqrt{c^2+3k^2} . 
\end{equation}
\end{prop}

The vanishing of the angular twist for 
the solitary waves (\ref{solitary1})--(\ref{solitary4}) 
and the kink (\ref{kink1})
means that the phase angle of $u(t,x)$ is the same (modulo $2\pi$) 
for both $x\rightarrow +\infty$ and $x\rightarrow -\infty$, 
as shown by the asymptotic behaviour 
(\ref{SSOLUTIONS1})--(\ref{SSOLUTIONS4}) and (\ref{SSOLUTIONS7}). 
For the kink (\ref{kink2}), 
the non-zero angular twist (\ref{Angulartwist2})
is produced by the finite net rotation in the phase angle of $u(t,x)$ 
between $x\rightarrow +\infty$ and $x\rightarrow -\infty$ 
as given by the asymptotic behaviour (\ref{SSOLUTIONS9}) 
when $\Theta \neq 0$. 
In contrast, 
the non-zero angular twist (\ref{Angulartwist3}) 
for the solitary wave (\ref{SWsolutions1}) arises from the linear phase
$k(x-(3c+8k^2)t)$. 

\begin{prop}
Peakon (\ref{SWsolutions3}) 
with speed $c>0$, phase angle $0\leq \phi<2\pi$, and frequency 
$-\infty<\epsilon c^{3/2}(1-3\sigma^2)/(\sigma^2-3)^{3/2} <\infty$
has the following conserved quantity:
\begin{equation}
\text{momentum}\qquad  
{\mathcal P}(c) =\sqrt{\frac{\sigma^2 -3}{c\sigma^2}} .
\end{equation}
\end{prop}

\begin{table}[!h]
\begin{changemargin}{-.2in}{.2in}
\begin{center}
\begin{tabular}{|c|c|c|c|c|c|c|}
\hline%\vspacebefore
\hfill&\hfill&\hfill & Galilean & phase-covariant & phase-covariant & angular \\
\hfill& momentum & energy & energy & mass & momentum & twist \\
\hline%\vspacebeforemore
$\begin{aligned}
&\text{Solitary}\\
&\text{wave (\ref{solitary1})
}\end{aligned}$
  & $\begin{aligned}&\Im\alpha=0\\&\beta=0\\& \Theta=0\end{aligned}$  
  & $\begin{aligned}&\Im\alpha=0\\&\beta=0\\& \Theta=0\end{aligned}$  
  & $\begin{aligned}&\Im\alpha=0\\&\beta=0\\& \Theta=0\end{aligned}$  
  &  - 
  &  - 
  & $\begin{aligned}&\Im\alpha=0\\&\beta=0\end{aligned}$  
\\[1.2ex]\hline%\vspacebeforemore
$\begin{aligned}
&\text{Solitary}\\
&\text{wave (\ref{solitary2})
}\end{aligned}$
  & - 
  & -
  & -
  & -  
  & -
  & $\begin{aligned}&\Im\alpha=0\\&\beta=0\end{aligned}$  
\\[1.2ex]\hline%\vspacebeforemore
$\begin{aligned}
&\text{Solitary}\\
&\text{wave (\ref{solitary3})
}\end{aligned}$
  & $\begin{aligned}&\Im\alpha=0\\&\Im\beta=0\\& \Theta=0\end{aligned}$  
  & $\begin{aligned}&\Im\alpha=0\\&\Im\beta=0\\& \Theta=0\end{aligned}$  
  & $\begin{aligned}&\Im\alpha=0\\&\Im\beta=0\\& \Theta=0\end{aligned}$  
  & $\begin{aligned}&\Re\alpha=2\Re\beta\\&\Im\alpha=\Im\beta=0\\& \Theta=0\end{aligned}$  
  & $\begin{aligned}&\Re\alpha=3\Re\beta\\&\Im\alpha=\Im\beta=0\\& \Theta=0\end{aligned}$  
  & $\begin{aligned}&\Im\alpha=0\\&\beta=0\end{aligned}$  
\\[1.2ex]\hline%\vspacebeforemore
$\begin{aligned}
&\text{Solitary}\\
&\text{wave (\ref{solitary4})
}\end{aligned}$
  & -
  & -
  & -
  & -
  & -
  & $\begin{aligned}&\Im\alpha=0\\&\beta=0\end{aligned}$  
\\[1.2ex]\hline%\vspacebeforemore
$\begin{aligned}
&\text{Solitary}\\
&\text{wave (\ref{SWsolutions1})
}\end{aligned}$
  & $\begin{aligned}&\Im\alpha=0\\&\beta=0\end{aligned}$  
  & $\begin{aligned}&\Im\alpha=0\\&\beta=0\end{aligned}$  
  & $\begin{aligned}&\Im\alpha=0\\&\beta=0\end{aligned}$  
  & -  
  & -
  & $\begin{aligned}&\Im\alpha=0\\&\beta=0\end{aligned}$  
\\[1.2ex]\hline%\vspacebeforemore
$\begin{aligned}
&\text{Peakon}\\
&\text{(\ref{SWsolutions3})
}\end{aligned}$
  & $\begin{aligned}&\Re\alpha=-\Re\beta\\&\Im\alpha=\Im\beta\end{aligned}$
  & -
  & -
  & -  
  & -
  & -
\\[1.2ex]
      \hline
      \end{tabular}
\smallskip
\end{center}
\caption{Conditions such that the integrated densities are conserved and finite}
\label{coeffconds} 
\end{changemargin}
\end{table}

\section{Summary}\label{summarize}

We have studied travelling wave solutions (\ref{travellingwave}) 
and conservation laws (\ref{2ndorderT})--(\ref{conslaw})
for a wide class of $U(1)$-invariant complex mKdV equations (\ref{mkdveqn}). 
This class contains the two known integrable generalizations of 
the ordinary (real) mKdV equation 
given by the Hirota mKdV equation and Sasa-Satsuma mKdV equation.
Four main results have been obtained. 

Firstly, 
in Theorems~\ref{solitarywavesolns} to \ref{kinksolns}, 
we respectively derive all solitary waves and all kinks of the complex form (\ref{travellingwave}), 
which have the asymptotic behaviour 
$u\rightarrow a$ as $|x|\rightarrow \infty$ 
and 
$u\rightarrow a \pm bf_0$ as $x\rightarrow \pm \infty$ 
where $a,b$ are complex constants and $f_0$ is a non-zero real constant. 
The solitary waves in Theorems ~\ref{solitarywavesolns} and~\ref{cuspsolns}
describe new complex generalizations of 
the well-known mKdV $\sech$ solution, 
and a new complex rational solution. 
The kinks in Theorem \ref{kinksolns} 
describe new complex generalizations of the familiar mKdV $\tanh$ solution.
Secondly,
in Theorems~\ref{linphasesolitarywaves} to \ref{linphasekinks},  
we respectively derive all solitary waves and kinks 
having a linear phase of the form (\ref{LINEARPHASE}),
with the asymptotic behaviour $|u|\rightarrow 0$ as $|x|\rightarrow \infty$ 
and $|u|\rightarrow f_0$ as $x\rightarrow \pm \infty$. 
These solutions include a new type of complex peakon 
whose phase has a jump discontinuity where the amplitude displays a cusp. 

Thirdly, 
in Theorem \ref{1storderT} we explicitly find 
all 1st order conserved densities (\ref{Cdensity}), 
which yield phase-invariant counterparts of 
the well-known mKdV conserved densities for 
momentum, energy, and Galilean energy, 
and a new conserved density which describes 
the angular twist of complex kink solutions. 
Fourthly, 
we explicitly find all conserved densities (\ref{2ndCdens}) of 
lowest order in higher $x$-derivatives. 
From Theorem \ref{TTS} the results of this classification establish that 
such higher-derivative conservation laws 
exist only in the two known integrable cases of complex mKdV equations.


\begin{thebibliography}{99}

\bibitem{RadLak} 
R. Radhakrishnan and M. Lakshmanan,
{\em Phys. Rev. E} 54 (1996) 2949--2955.

\bibitem{GilHieNimOht}
C. Gilson, J. Hietarinta, J. Nimmo, Y. Ohta, 
{\em Phys. Rev. E} 68 (2003) 016614 (10 pages). 

\bibitem{Hir1973}
R. Hirota, 
{\em J. Math. Phys.} 14 (1973) 805--809.

\bibitem{SasSat1991}
N. Sasa and J. Satsuma,
{\em J. Phys. Soc. Jpn.} 60 (1991) 409--417.

\bibitem{GaoTan}
Y. Gao and X.-Y. Tang,
{\em Commun. Theor. Phys.} 48 (2007) 961--970. 

\bibitem{Hir1972}
R. Hirota, 
{\em J. Phys. Soc. Jpn.} 22 (1972) 1456--1458.

\bibitem{MiuGarKru}
R.M. Miura, C.S. Gardner, M.S. Kruskal,
{\em J. Math. Phys.} 9 (1968) 1204--1209.

\bibitem{1stbook}
G. Bluman and S.C. Anco,
{\it Symmetry and Integration Methods for Differential Equations},
Springer Applied Mathematics Series V.154, 2002.

\bibitem{AncTchWil}
S.C. Anco, N. Tchegoum Ngatat, M. Willoughby,
{\em Physica D} 240 (2011) 1378--1394.

\bibitem{Olvbook}
P. Olver,
{\it Applications of Lie Groups to Differential Equations},
Springer-Verlag, New York, 1986.

\bibitem{2ndbook}
G. Bluman, A. Cheviakov, S.C. Anco, 
{\it Applications of Symmetry Methods to Partial Differential Equations}, 
Springer Applied Mathematics Series V.168, 2010.

\bibitem{AncBlu2002a}
S.C. Anco and G. Bluman, 
%Direct construction method for conservation laws of partial differential equations.~I. Examples of conservation law classifications,
{\it Euro. J. Appl. Math.} 13 (2002), 545--566.
%(math-ph/0108023)

\bibitem{AncBlu2002b}
S.C. Anco and G. Bluman, 
%Direct construction method for conservation laws of partial differential equations.~II. General treatment,
{\it Euro. J. Appl. Math.} 13 (2002), 567--585.
%(math-ph/0108024)

\bibitem{crack}
T. Wolf,
Applications of CRACK in the Classification of Integrable Systems,
in CRM Proceedings and Lecture Notes, 37 (2004) 283--300.
http://lie.math.brocku.ca/crack/demo/

\bibitem{conlaw}
T. Wolf,
Crack, LiePDE, ApplySym and ConLaw, section 4.3.5 in:
Grabmeier J., Kaltofen E. and Weispfenning~V. (Eds.), 
{\it Computer Algebra Handbook}, Springer, 2002, 465--468.



\end{thebibliography}
\end{document}